\documentclass{piner}
\usepackage{rotating}
\begin{document}

\shorttitle{Limb-Brightening in Markarian 501}

\title{Significant Limb-Brightening in the Inner Parsec of Markarian 501}

\author{B.~Glenn~Piner\altaffilmark{1}, Niraj~Pant\altaffilmark{1}, Philip~G.~Edwards\altaffilmark{2},
and Kaj~Wiik\altaffilmark{3}}

\altaffiltext{1}{Department of Physics and Astronomy, Whittier College,
13406 E. Philadelphia Street, Whittier, CA 90608; gpiner@whittier.edu}

\altaffiltext{2}{CSIRO Australia Telescope National Facility, 
Locked Bag 194, Narrabri NSW 2390, Australia; Philip.Edwards@csiro.au} 

\altaffiltext{3}{Tuorla Observatory, University of Turku, V\"{a}is\"{a}l\"{a}ntie 20, 
FI-21500 Piikki\"{o}, Finland}

\begin{abstract}
We present three 43~GHz images and a single 86~GHz image of Markarian 501
from VLBA observations in 2005. The 86~GHz image shows a partially resolved core with a flux density of
about 200~mJy and a size of about 300 Schwarzschild radii, similar to recent results
by Giroletti et al.
Extreme limb-brightening is found in the inner parsec of the jet in the 43~GHz images,
providing strong observational support for a `spine-layer' structure at this distance from the core.
The jet is well resolved transverse to its axis, allowing Gaussian model components
to be fit to each limb of the jet. The spine-layer brightness ratio and relative sizes,
the jet opening angle, and a tentative detection of superluminal motion in the layer
are all discussed.
\end{abstract}


\section{Introduction}
\label{intro}
Markarian 501 is among the best known and most prominent of the TeV blazars.
It was the second such source to be discovered (Quinn et al. 1996), and during
flaring episodes in 2005 it displayed TeV variability on timescales
of several minutes (Albert et al. 2007). The remarkably fast variability
found for this source (and for PKS~2155$-$304) may require very high Lorentz
factors ($\Gamma>50$) in the gamma-ray production region   
(Begelman et al. 2008).

In contrast to the TeV observations, VLBI studies of jet kinematics in the TeV blazars
(including Mrk 501, Edwards \& Piner [2002]) have suggested much lower Lorentz factors ($\Gamma\sim3$) in their 
parsec-scale VLBI jets (Piner et al. 2008).
A number of explanations for this discrepancy have been discussed in the literature,
with one possibility being that the jet is structured transverse to its axis
into a fast `spine' that produces the high-energy emission, and a slower `layer'
responsible for the radio emission (e.g., Ghisellini et al. 2005).
Transverse jet structures are also expected on theoretical grounds (e.g., Zakamska et al. 2008).
Limb-brightened VLBI jets that confirm such transverse structures
have been seen in M87 (Ly et al. 2007; Kovalev et al. 2007) and in Mrk~501 itself 
at frequencies below 22~GHz (Giroletti et al. 2004, hereafter G04), and tentatively at high frequencies
(Giroletti et al. 2008, hereafter G08).

In this paper, we report the detection of significant limb-brightening in the inner parsec
of Mrk~501, based on a sequence of three 43~GHz VLBA images from 2005.
We use cosmological parameters $H_{0}=71$ km s$^{-1}$
Mpc$^{-1}$, $\Omega_{m}=0.27$, and $\Omega_{\Lambda}=0.73$. 
At the redshift of Mrk~501 (0.034), an
angular separation of 1 milliarcsecond (mas) corresponds to a projected linear
distance of 0.66~pc. We assume a black hole mass of $10^{9} M_{\odot}$ (Wagner 2008),
consistent with that used by G08. 

\section{Observations and Data Reduction}
\label{obs}
We observed Mrk~421 and Mrk~501 at three epochs during 2005 with 8 elements (without Saint Croix and Hancock) of the
National Radio Astronomy Observatory's Very Long Baseline Array (VLBA),
under observation code BP112. Results for Mrk~421 will be reported elsewhere.
At each epoch, Mrk~501 was observed for approximately 90 minutes at 86~GHz
and 30 minutes at 43~GHz,
at a high data rate of 512~Mbps. 
Each 86~GHz scan on Mrk~501 was preceded by a pointing scan at 43~GHz on the bright SiO maser U~Her.
The AIPS software package was used for calibration
and fringe-fitting of the correlated visibilities, and the
final CLEAN images and self-calibrated visibilities were produced using
the Difmap software package.

Mrk~501 is near the VLBA's detection limit at 86~GHz, even at our high data rate
of 512 Mbps, and fringes were detected
at 86~GHz only at the first epoch, when Mrk~501 was at its brightest in the 43~GHz
images. Successful 43~GHz images were made at all three epochs.
The single 86~GHz image and the three 43~GHz images are shown in Figure~1.
All of the images have an rms noise level
within a factor of two of the expected thermal noise limits.

\begin{figure*}[!t]
\begin{center}
\includegraphics[scale=0.57]{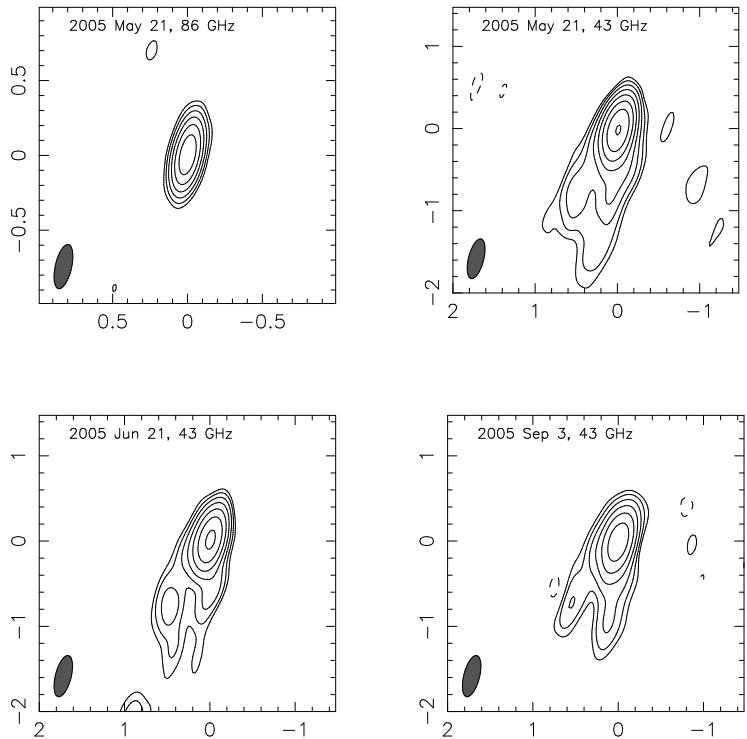}
\caption{VLBA images of Markarian 501 at 86 and 43~GHz.
Axes are labeled in milliarcseconds (mas).
The 86~GHz image has a beam size of 0.30 by 0.11~mas at position 
angle $-13\arcdeg$, a peak flux density of 221~mJy beam$^{-1}$,
and a lowest contour of 4.0~mJy beam$^{-1}$.
The 43~GHz images are all restored with the same beam size of
0.50 by 0.19~mas at position angle $-14\arcdeg$ (the average 
beam size from the three epochs). 
Peak flux densities in the 43~GHz images are 278, 221, and 179~mJy beam$^{-1}$.
The lowest contours in the 43~GHz images are 2.1, 3.0, and 3.4~mJy beam$^{-1}$.
Lowest contours are all set to three times the rms noise level.
Subsequent contours increase by factors of two.
The feature evident at $\sim$2.5~mas from the core in the 2005 Jun 21 image may be associated 
with the component C3 described in Edwards \& Piner (2002), and is only detected at this epoch.}
\end{center}
\end{figure*}

The 86~GHz image shows a compact core component with a peak flux density of
about 200~mJy beam$^{-1}$, and no detectable jet emission.
Because the source was at the threshold of detection at 86~GHz, we applied the 
test based on the averaging time of the selfcal solutions described 
by Mart\'{i}-Vidal \& Marcaide (2008),
in order to make sure that it was not a spurious source generated by self-calibration
of noisy data, and the real nature of the source was verified.

The three 43~GHz images show a limb-brightened jet extending out to
about 1~mas (0.66~pc projected) from the core, that is well resolved in the transverse
direction (about five beams across the full-width of the jet).
The limb-brightening is very prominent in these 43~GHz images, particularly
at the third epoch when no emission is detected in the center of the jet at all.
The final 43~GHz image is close in time to the 86~GHz image of G08, where they
report the detection of a tentative jet knot at $(r,\theta)=(0.72\;{\rm mas}, 172\arcdeg)$.  
We identify this tentative jet knot with the local peak in the western limb on
the final 43~GHz image, suggesting that the G08 `tentative knot' is a peak in an
underlying distribution, and not an isolated component.

We fit elliptical Gaussian components to the visibilities
associated with the four images in Figure~1, using the modelfit task in Difmap.
These elliptical Gaussian model fits are tabulated in Table~\ref{mfittab}.
The 86~GHz data were fit by a single elliptical Gaussian.
For all of the 43~GHz epochs,
an elliptical Gaussian was required for the core and for each limb of the limb-brightened jet
at similar distances from the core. Each limb was well-fit by a long, thin, elliptical Gaussian
oriented approximately along the jet direction.
These elliptical Gaussians are labeled
`eastern limb' and `western limb' in the model fits in Table~\ref{mfittab}. 
Recall that east and west are reversed on the images relative to their standard directions.
That region of the jet corresponds to the region that was fit by a single stationary Gaussian
component (called C4) in our earlier work at lower resolution (Edwards \& Piner 2002).
These observations then confirm what was already noted about this source by G04; that
at higher resolutions the jet has a limb-brightened structure that
cannot be well represented by a single Gaussian component.

\begin{table*}[!t]
\caption{Elliptical Gaussian Models} 
\vspace{-0.1in}
\begin{center}
\label{mfittab}
{\scriptsize \begin{tabular}{l c c c c c c c r} \colrule \colrule \\ [-5pt]
& Frequency & & $S$ & $r$ &
PA & $a$ & & \multicolumn{1}{c}{$\phi$} \\
\multicolumn{1}{c}{Epoch} & (GHz) & Component & (mJy) & (mas) &
(deg) & (mas) & $b/a$ & (deg) \\
\multicolumn{1}{c}{(1)} & (2) & (3) & (4) & (5) &
(6) & (7) & (8) & \multicolumn{1}{c}{(9)} \\ \colrule \\ [-5pt]
2005 May 21 & 86 & Core         & 231 & ...  & ...   & 0.045 & 0.46 & 23.3    \\ 
2005 May 21 & 43 & Core         & 344 & ...  & ...   & 0.137 & 0.81 & $-$89.1 \\
            &    & Eastern limb & 60  & 0.67 & 143.5 & 1.128 & 0.21 & $-$24.3 \\
            &    & Western limb & 58  & 0.68 & 178.1 & 0.832 & 0.26 & $-$24.6 \\
2005 Jun 21 & 43 & Core         & 269 & ...  & ...   & 0.139 & 0.77 & $-$68.7 \\
            &    & Eastern limb & 62  & 0.93 & 152.8 & 1.778 & 0.12 & $-$11.1 \\
            &    & Western limb & 28  & 0.67 & 178.5 & 0.511 & 0.19 & $-$23.3 \\
2005 Sep 3  & 43 & Core         & 255 & ...  & ...   & 0.228 & 0.66 & $-$35.5 \\
            &    & Eastern limb & 39  & 0.62 & 137.0 & 1.146 & 0.07 & $-$25.1 \\
            &    & Western limb & 48  & 0.76 & 175.1 & 0.645 & 0.36 & $-$17.9 \\ \colrule
\end{tabular}}
\end{center}
NOTES.--- Col. (4): Flux density in millijanskys.
Col. (5) and (6): Polar coordinates of the
center of the component relative to the core.
Position angle is measured from north through east.
Col. (7), (8), and (9): FWHM of the major axis,
axial ratio, and position angle of the major axis.
\end{table*}

\section {Results and Discussion}
\label{results}

\subsection{VLBI Core}
The model fit to the VLBI core at 86~GHz is a 231~mJy ellipse of size
0.045 by 0.021~mas.
The major axis is resolved, but not the minor axis --- the upper limit on the minor
axis size is $<0.030$~mas, corresponding to a brightness temperature $>3\times10^{10}$~K.  
The model fit constrains the most compact radio emission in Mrk~501 to come from a size
less than 300x200 Schwarzschild radii, the same size upper limit that was found by G08.
Comparing to the 86~GHz image of this source from October 2005 by G08, we note that G08 used the
Global Millimeter VLBI Array (GMVA), so their beam size is about a factor of 2-3
times smaller. They resolve the central region into three components, a core and two
components that may represent the broad base of the limb-brightened jet; these
features are likely blended with the core in our lower resolution image.
G08 find a total flux density of 110~mJy in the core region, less than our measured
core flux density, but Mrk~501 faded between May and September 2005 at 43~GHz, and likely
did so at 86~GHz as well.

The spectral index of the core between 43 and 86~GHz (with sign convention $S\propto\nu^{+\alpha}$), 
measured using the model-fit core components from May 2005,
is $-0.6\pm0.2$, assuming 10\% and 20\% accuracies for the amplitude calibrations 
at 43 and 86~GHz, respectively (from correction factors given by the gscale command in Difmap). 
This is consistent with the value of $-0.5$ measured
for the high-frequency spectral index by G04 and G08.  
The flux density of the VLBI core drops at 43~GHz over our three epochs, from 344 to
255~mJy (0.10 probability of no variation, assuming 10\% amplitude errors).
A major gamma-ray flare was detected just nine days after our second epoch (Albert et al. 2007),
but this does not seem to have manifested in a VLBI core flare, at least not at any
of our epochs. A similar fading of the VLBI core during a time of high TeV activity 
was seen by Charlot et al. (2006) in Mrk~421.

\subsection{Limb-Brightened Jet}
The major result of this work is the detection of the prominent limb-brightening
of the jet in its inner parsec in the three 43~GHz images,
definitively confirming a spine-layer structure in this region of the jet.
The limb-brightening is prominent enough 
that the model fits yield separate elliptical Gaussian components for
each limb of the jet, allowing us to make measurements of the layer
properties free from beam convolution effects.

Figure~2 shows the transverse brightness across the jet from the 
CLEAN image from the third epoch, at a distance of 0.75~mas (0.5~pc projected) from the core.
From such transverse brightness plots, or from the model fits, we can set lower limits on the
layer/spine brightness ratio.
The layer/spine brightness ratio in Figure~2 is about 5:1, and although the measured ratio
changes with epoch and distance from the core, this is close to its median value of 6:1.
This ratio is a lower limit, because the images used to obtain
the transverse plots have been convolved with the CLEAN beam.
From the model fits, we place a 3$\sigma$ lower limit on the layer/spine brightness
ratio of $>$10:1, by comparing the peak surface brightness of the limb model components
in Table~\ref{mfittab} to three times the rms noise level.

\begin{figure*}[!t]
\begin{center}
\includegraphics[scale=0.50]{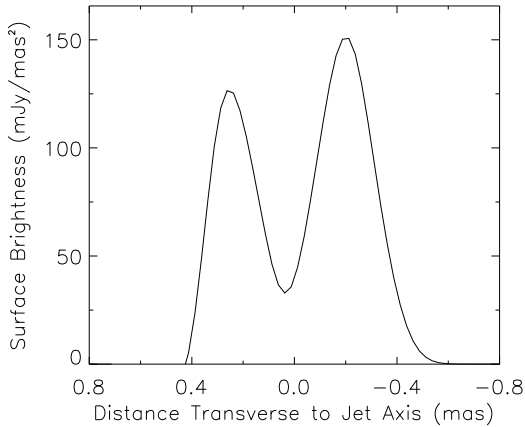}
\caption{Transverse brightness across the jet from the 
CLEAN image from 2005 Sep 3, at a distance of 0.75~mas (0.5~pc projected) from the core.
The CLEAN components representing the core have been subtracted to avoid contamination
from core emission.}
\end{center}
\end{figure*}

These measured ratios can be compared to that predicted by the geometric model for
limb-brightening in this source by G04.
In that model, limb brightening is due to the slower layer acquiring a higher
Doppler factor than the fast spine after a bend of the jet
away from the line-of-sight. In this jet region, G04 estimate $\delta_{layer}\sim5$ and
$\delta_{spine}\sim2$. Using the beaming exponent for a continuous jet, ($2-\alpha$), where
$\alpha=-0.5$, and assuming equal unbeamed emission for the spine and layer, the G04
model predicts a brightness ratio of 10, just consistent with our observed upper limit.
More sophisticated models for transverse structures (e.g., Ghisellini et al. 2005;
Zakamska et al. 2008) do not invoke jet bending, and instead rely on
different emissivities in the spine and layer (see, e.g., Figure~4 in Zakamska et al. 2008).

A series of transverse brightness plots like that in Figure~2 may be used to
produce a plot of the peak brightness in each limb as a function of distance from the core.
Features in the limbs may then potentially be followed in time to yield an
estimate of the apparent pattern speed in the layer.
Figure~3 shows the peak brightness in the eastern and western limbs vs. distance
from the core for the second epoch.
Note the different locations of the peak emission in the western and eastern limbs.
Over the three epochs,
the location of the peak in the eastern limb remains stationary, but the location of the peak in the western
limb changes from $r=0.425$ to 0.50 to 0.825~mas from the core 
(this is also visible in the CLEAN images in Figure~1).
If interpreted as a pattern speed, this yields an apparent speed of $3.3\pm0.3~c$ in the western limb.
Such a speed is consistent with the spine-layer model for this source proposed by G04.
They model the layer at this distance from the core with a Lorentz factor of 3 and an angle to the line-of-sight
of $15\arcdeg$, which would yield an apparent speed of $3~c$.
\footnote{In this case, the stationary feature in the eastern limb may simply be one
of the stationary patterns common to blazar jets, which often show moving and stationary
patterns in the same source, e.g., Jorstad et al. (2001).}
This apparent speed measurement should be regarded as tentative; further 43 GHz observations
of Mrk~501 to be taken this year should provide more data on the time evolution of the layer structure
(for example, will helical motion be observed for layer features if monitored over a
longer time range?).

\begin{figure*}
\begin{center}
\includegraphics[scale=0.50]{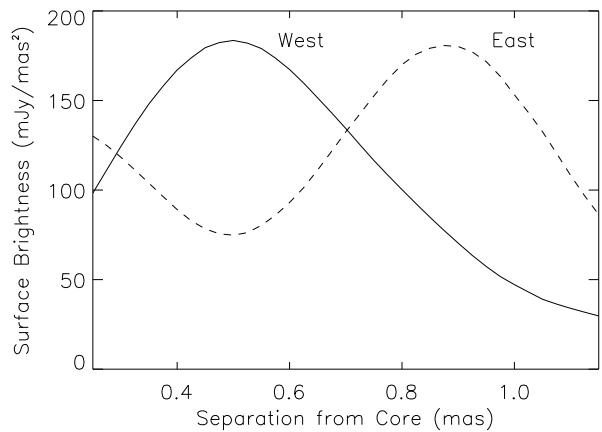}
\caption{Peak brightness in each limb as a function of distance from the core in the CLEAN image from
2005 Jun 21. The peak brightness in the western limb is plotted
as a solid line, the peak brightness in the eastern limb is plotted as a dashed line.}
\end{center}
\end{figure*}

Because the jet is transversely resolved, we can measure the jet
width and apparent opening angle as a function of distance from the core.  
The apparent opening half angle of the jet in the G08 86~GHz image is very wide ($\sim40\arcdeg$)
at about 0.1~mas from the core,
but this opening angle does not persist --- the plots of jet width vs. distance from the
core shown by G04 show a much narrower opening angle ($\sim5\arcdeg$) at their first
point at about 4~mas.
Our 43~GHz data show strong collimation in an approximately cylindrical jet in the
intermediate region from about 0.3 to 1~mas (0.2 to 0.7~pc projected distance).
We calculate the jet width and opening angle from the model fits in Table~\ref{mfittab},
with the width at each radius equal to the separation of the major axes of the
ellipses fit to the two limbs. The opening angle vs. distance calculated in this way is
shown in Figure~4 for the first epoch. All epochs give similar results: the jet width
is approximately constant and the opening angle decreases approximately as $1/r$.
Using transverse brightness plots (like Figure~2) to measure the same quantities
gives qualitatively identical results, but with larger values for the width
and opening angle, because those plots have been convolved with the beam.

\begin{figure*}[!t]
\begin{center}
\includegraphics[scale=0.59]{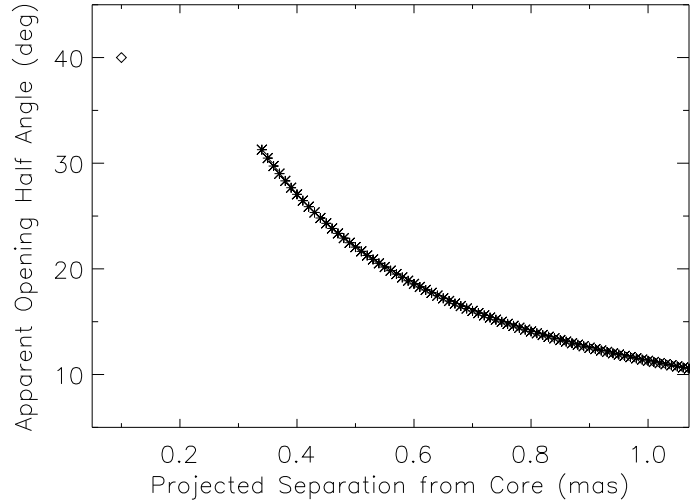}
\caption{Apparent half-opening angle vs. projected core distance for 2005 May 21.
The opening angle is calculated at each radius from the model-fits in Table~\ref{mfittab}.
The diamond shows the approximate opening angle measured by the G08 86~GHz observation.}
\end{center}
\end{figure*}

We used the DIFWRAP package for model component error analysis (Lovell 2000)
to investigate whether or not the ellipses fit to the limbs of the jet
are resolved in their minor axis direction, or whether ellipses with zero
axial ratio also provide an adequate fit to the visibilities.
The DIFWRAP package allows for visual inspection of the fits to the visibilities
as the parameter space of the model is methodically searched. We find that,
at the epoch with the highest quality data (the first epoch),
the elliptical Gaussians fit to the limbs in Table~\ref{mfittab} are resolved 
in their minor axis direction (i.e., an ellipse
of zero axial ratio does not fit the visibilities),
and this allows us to estimate the relative sizes of the layer and spine. 
In a simple model where brightness is proportional to the path length
of the line-of-sight through the layer, the peak brightness occurs at the edge
of the spine, and half the separation of the major axes of the ellipses
gives the radius of the spine ($r_{spine}=0.2$~mas, or 0.13~pc). The semi-minor axes of the ellipses then
give the points at which the line-of-sight path length through the layer has decreased
by a factor of two. Using these geometric constraints, and the first epoch model in Table~\ref{mfittab},
we compute the outer radius of the layer. We find that the layer occupies a volume
about twice that of the spine ($r_{layer}=1.7r_{spine}$). 
These measurements are difficult because they rely on resolving the limb in
the transverse direction, and hopefully will be confirmed by additional study.

\section{Conclusions}
The high resolution of the 43~GHz VLBA observations, combined with the sensitivity
of their 512 Mbps bandwidth, has enabled us to resolve the
spine-layer structure of the jet in the inner parsec of Mrk~501 with unprecedented
clarity. For the first time, our data resolve the transverse structure
well enough that separate Gaussian models can be fit to each limb, which is
crucial for measurements of the structure deconvolved from the observing beam.
Below we enumerate the major results on the core and limb-brightened jet obtained in this Letter:
\begin{enumerate}
\item{The $3\sigma$ lower limit on the layer/spine brightness ratio measured from the
Gaussian model fits at 0.5 pc (projected) from the core is $>$10:1.}
\item{A brightness peak in the layer appears to move outward with an apparent pattern
speed of $3.3\pm0.3~c$.}
\item{The jet is approximately cylindrical from 0.2 to 0.7 pc (projected) from the core,
and the apparent opening half angle decreases from about $30\arcdeg$ to about $10\arcdeg$ over this
region (Figure~4).}
\item{An estimate of the relative sizes of the layer and spine (from an epoch where the minor axes
of the ellipses fit to the layer are resolved) shows that the volume of the layer is about
twice that of the spine.}
\item{The detection of the 86~GHz core constrains the most
compact radio emission in Mrk~501 to come from a size
less than 300x200 Schwarzschild radii, the same size upper limit found by G08.}
\end{enumerate}
These results can all aid in constraining fundamental properties of the
limb-brightened structure; like the Lorentz factor, viewing angle, and Doppler factor of
the spine and layer, which are important for modeling
the radio to TeV gamma-ray emission from the two transverse regions of the
jet in this source.
Further observations currently in the VLBA's dynamic queue should reveal more about the
nature of the inner jet structure of Mrk~501, during the first year of Fermi (GLAST) observations.

\vspace{-0.1in}
\acknowledgments
We acknowledge helpful comments from the referee that improved
the quality of the paper.
The National Radio Astronomy Observatory is a facility of the National
Science Foundation operated under cooperative agreement by Associated
Universities, Inc.  
This work was
supported by the National Science Foundation under Grant 0707523.

{\it Facilities:} \facility{VLBA ()}

\end{document}